# Internal quantum efficiency of AlGaN/AlN quantum dot superlattices for electron-pumped ultraviolet sources

A. Harikumar[1], F. Donatini[2], C. Bougerol[2], E. Bellet-Amalric[1], Q.-M. Thai[3], C. Dujardin[3], I. Dimkou[4], S. T. Purcell[3], and E. Monroy[1]

[1] Univ. Grenoble-Alpes, CEA-IRIG-PHELIQS, 17 av. des Martyrs, 38000 Grenoble, France.
[2] Univ. Grenoble-Alpes, CNRS Institut Néel, 25 av. des Martyrs, 38000 Grenoble, France.
[3] Univ. Lyon, Université Claude Bernard Lyon 1, CNRS, Institut Lumière Matière, 69622 Lyon, France
[4] Univ. Grenoble-Alpes, CEA, LETI, 17 av. des Martyrs, 38000 Grenoble, France.

## Abstract

In this paper, we describe the growth and characterization of ≈ 530-nm-thick superlattices (100 periods) of $Al_xGa_{1-x}N$/AlN ($0 \leq x \leq 0.1$) Stranski-Krastanov quantum dots for application as the active region of electron-beam pumped ultraviolet lamps. Highly dense (>$10^{11}$ cm$^{-2}$) quantum dot layers are deposited by molecular beam epitaxy, and we explore the effect of the III/V ratio during the growth process on their optical performance. The study considers structures emitting in the 244-335 nm range at room temperature, with a relative linewidth in the 6-11% range, mainly due to the QD diameter dispersion inherent in self-assembled growth. Under electron pumping, the emission efficiency remains constant for acceleration voltages below ≈ 9 kV. The correlation of this threshold with the total thickness of the superlattice and the penetration depth of the electron beam confirms the homogeneity of the nanostructures along the growth axis. Below the threshold, the emission intensity scales linearly with the injected current. The internal quantum efficiency is characterized at low injection, which reveals the material properties in terms of non-radiative processes, and high injection, which emulates carrier injection in operation conditions. In quantum dots synthesized with III/V ratio < 0.75, the internal quantum efficiency remains around 50% from low injection to pumping power densities as high as 200 kW/cm$^2$, being the first kind of nanostructures that present such stable behaviour.

# 1. Introduction

Semiconductors emitting in the deep ultraviolet (UV) range are in high demand for various applications in disinfection, water purification and related fields. This is due to the fact that most viruses and bacteria experience photochemical changes to the nucleic acids (for example, by the formation of dimers in deoxyribonucleic or ribonucleic acids) when exposed to UV-C radiation, which damages their ability to reproduce and therefore makes them inactive [1]. Mercury lamps are being used at present for this purpose; however, they are highly toxic and present low shelf life. Without question there is a strong need to replace them. Light emitting diodes (LEDs) based on AlGaN semiconductors are being pushed as a replacement [2,3] as their emission is spectrally tunable in the entire UV range, they are non-toxic, eco-friendly and exhibit long lifetime and fast switching capabilities. However, despite there being a plethora of research in III-V LEDs [1,4,5], their performances are still not at the level of their arc lamp counterparts. Laboratories report record LED external quantum efficiencies (EQE) close to 20% at 275 nm [6], yet commercial devices exhibit efficiencies lower than 1%. The limitations are mostly attributed to doping, asymmetric carrier transport and metal contacting.

To circumvent these issues we propose electron-beam pumped UV lamps based on $Al_xGa_{1-x}N$/AlN quantum dot (QD) superlattices as active region. Devices based on this concept consist of a vacuum tube with a cold cathode (e.g. a carbon nanotube grid) which injects electrons in the semiconductor chip and generates electron hole pairs in the active region by impact ionization [7–9]. In principle, they can be miniaturized to the millimetre size. No longer needing p-type doping or contacts, we already tackle a few major problems concerning LEDs, namely the asymmetric carrier mobility, ohmic-contacting, and issues related to doping ionization energy. Electron-pumped emission from GaN-based quantum wells has been reported [8,10–14], but the performance was mostly limited by the light extraction efficiency. Electron-pumped lasing around 350 nm has also been demonstrated [15,16].

Carrier localization in potential fluctuations leads to an enhancement of the emission yield at room temperature [17–21], which has motivated research on QDs for the development of high efficiency UV emitters [22–27]. In particular, plasma-assisted molecular beam epitaxy (MBE) is known to produce layers of close-packed Stranski-Krastanov (SK) GaN QDs with high internal quantum efficiency [17,23,24]. We have previously reported $Al_xGa_{1-x}N$ QDs emitting at wavelengths as short as 235 nm at room temperature with internal quantum efficiency IQE ≈ 30% [23]. We have also demonstrated IQE higher than 50% in the 276-296 nm spectral range, keeping in mind their high sensitivity to the amount of $Al_xGa_{1-x}N$ in the QD layer [22].

In this paper, we explore the effect of the III/V ratio during the deposition of $Al_xGa_{1-x}N$/AlN Stranski-Krastanov (SK) QDs on their optical properties, in view of their application as the active region for electron-beam pumped UV emitters. Their IQE is characterized as a function of the optical pumping power, covering low excitation densities, which reveals the material properties in terms of nonradiative processes, and high excitation densities, which emulates carrier injection in operation conditions. In the case of electron-beam pumping, the variation of the efficiency as a function of the acceleration voltage and injection current is also discussed.

## 2. Methods

The periodicity of the structures was analysed by X-ray diffraction (XRD) in a Rigaku SmartLab diffractometer using a 2-bounce Ge (220) monochromator and a long plate collimator of 0.228° for the secondary optics. The morphology of the QDs was analysed by atomic force microscopy (AFM) using a Bruker ICON SPM system operated in the tapping mode using TEXPA-V2 probes. Additional structural studies were conducted using high-resolution transmission electron microscopy (HRTEM) and high-angle annular dark-field (HAADF) scanning transmission electron microscopy (STEM) performed on a FEI Tecnai microscope operated at 200 kV.

Photoluminescence (PL) measurements under continuous-wave excitation were obtained by pumping with a frequency-doubled solid-state laser ($\lambda$ = 244 nm), with an optical power of 100 µW focused on a spot with a diameter of ≈ 100 µm. PL measurements under pulsed excitation used an Nd-YAG laser (266 nm, 2 ns pulses, repetition rate of 8 kHz). In both cases, samples were mounted on a cold-finger cryostat, and the PL emission was collected by a Jobin Yvon HR460 monochromator equipped with a UV-enhanced charge-coupled device (CCD) camera.

Cathodoluminescence (CL) experiments were carried out using a FEI Inspect F50 field-emission SEM equipped with a low-temperature Gatan stage to cool the sample down to 6 K, and with an iHR550 spectrometer fitted with a Andor Technology Newton DU940 BU2 spectroscopic CCD camera. The beam spot diameter was ≈ 10 nm on the focal point, the accelerating voltage was varied from 2 to 20 kV, and the electron beam current was kept below 150 pA. Additional CL experiments were performed using a Kimball Physics EGPS-3212 electron gun operated in direct current mode, under normal incidence, with a beam spot diameter of 4±1 mm. The acceleration voltage was in the range of 3 to 10 kV, injecting up to 800 µA of current. The CL emission arrived to an ANDOR ME-OPT-0007 UV-NIR light collector coupled with an ANDOR Shamrock500i spectrograph connected to an electron-multiplying CCD Newton 970 from ANDOR operated in conventional mode.

The electronic structure of the QDs was modelled in three dimensions (3D) using the Nextnano[3] 8-band **k·p** Schrödinger-Poisson equation solver [28] with the material parameters described in ref. [29]. For the $Al_xGa_{1-x}N$ alloys, all the bowing parameters were set to zero. The simulated structure consisted of 10 layers of $Al_xGa_{1-x}N$ QDs embedded in AlN. The period of the structure (QD layer + AlN barrier) was fixed to 5.3 nm. In each QD layer, 7 QDs were defined as a hexagonal truncated pyramid with {10-13} facets [28], distributed in-plane in a hexagonal close-packed configuration, and connected by a wetting layer. The 3D strain distribution was obtained by minimization of the elastic energy through the application of periodic boundary conditions along the <1-100> and <11-20> in-plane directions. For the calculation of the band diagram, the spontaneous and piezoelectric polarization and the band gap deformation potentials were taken into account. It was assumed that the lattice was at room temperature, and the temperature dependence of the band gap follows Varshni equation. The quantum confined levels for electrons and holes and their associated square wavefunctions, $|\Psi(r)|^2$, were calculated in the QD located in the centre of the structure.

## 3. Material growth

In order to explore the advantages of using $Al_xGa_{1-x}N$/AlN Stranski-Krastanov (SK) QDs as the active region for electron-pumped UV emitters, we synthesized samples consisting of a stack of 100 layers of self-assembled $Al_xGa_{1-x}N$ QDs with x = 0 or 0.1, separated by 4 nm of AlN as barrier. Such QD superlattices were grown using plasma-assisted MBE on 1-µm-thick (0001)-oriented AlN-on-sapphire templates at a temperature $T_S$ = 720°C. The growth process was monitored by reflection high-energy electron diffraction (RHEED). The active nitrogen flux, $\Phi_N$, was tuned to achieve a growth rate $v_G = \Phi_N = 0.52$ monolayers/s (ML/s) under metal rich conditions. Note that 1 monolayer (ML) of AlN or GaN is ≈ 0.25 nm. The $Al_xGa_{1-x}N$ dots were grown using N-rich conditions, which are known to lead to a high density of small QDs [17,18,22,29]. We explored the effect of varying the Ga flux (i.e. the gallium-to-nitrogen flux ratio) in the range of $\Phi_{Ga}$ = 0.149-0.441 ML/s ($\Phi_{Ga}/\Phi_N$ = 0.29-0.85), keeping the Ga deposition time constant (12 s). In the case of ternary $Al_xGa_{1-x}N$ QDs, we added a flux of aluminium $\Phi_{Al}$ so that the targeted Al mole fraction in the dots is x = $\Phi_{Al}/(\Phi_{Al} + \Phi_{Ga})$. The deposition of the QDs was followed by a growth interruption of 15 s. At the end of this process, a spotty RHEED pattern confirmed the presence of QDs. To favour the charge evacuation during the electron pumping process, the QDs were doped n-type with [Si] = $5\times10^{18}$ cm$^{-3}$ (value estimated from Hall effect measurements using the Van der Pauw method on planar Si-doped GaN layers). The presence of silicon during the growth process does not have any effect on the growth kinetics or in the resulting QD shape/density [29]. The AlN sections were grown under Al-rich conditions ($\Phi_{Al}/\Phi_N$ = 1.1), followed by a growth interruption under nitrogen to consume the accumulated Al excess. At the end of the growth of each AlN barrier, the RHEED pattern showed the straight lines characteristic of a planar surface. The growth process was sharply interrupted (the precursor fluxes were shuttered, and the substrate was rapidly cooled down) after the deposition of the last QD layer, to enable AFM characterization of the QD shape and density. A schematic description of the general structure is depicted in figure 1(a) and a summary of the samples under study is presented in table 1.

## 4. Results and discussion

The structural properties of the QD superlattices were characterized by XRD. In a reciprocal space map around the (10-15) reflection, depicted in Figure 1(b) for sample

SG5, we can identify the reflection from the AlN substrate and the satellites of the QD superlattice (SL), which are vertically aligned. The satellites and the AlN reflection present the same in-plane reciprocal vector, $Q_x = -2/(a\sqrt{3}) = -3.710$ nm$^{-1}$, where $a$ is the in-plane lattice parameter of AlN. This is an indication of pseudomorphic growth, where the mismatch stress is elastically released. Note that for this reflection, the out of plane reciprocal vector is $Q_z = 1/(5c)$, where $c$ is the out of plane lattice parameter. The period of the structures was extracted from the inter-satellite distance in a θ–2θ scan around the (0002) reflection of AlN, as shown in figure 1(c) for 3 Al$_{0.1}$Ga$_{0.9}$N/AlN QD samples grown with different metal-to-nitrogen flux ratio (SA1, SA2, and SA3), and GaN/AlN QD superlattice (SG5). The large number of satellites reveals the high quality of the samples even after the growth of 100 QD layers. In the figure, the experimental results are compared with a theoretical calculation using the Rigaku SmartLab Studio II software. The calculation assumes a period of 5.43 nm and that average out of plane lattice parameter of the QD superlattice is that of Al$_{0.90}$Ga$_{0.10}$N fully strained on AlN. In all the samples, the period was 5.3±0.3 nm, in good agreement with the nominal growth parameters. Therefore, the total thickness of the QD stack is ≈ 530 nm.

The morphology of the topmost QD layers was analysed by AFM, with the result presented in figure 2 for samples SG3 (GaN/AlN QDs) and SA1 (Al$_{0.1}$Ga$_{0.9}$N/AlN QDs). In such images, the QD density is 3.3±0.4×10$^{11}$ cm$^{-2}$ and 4.6±0.5×10$^{11}$ cm$^{-2}$, respectively, and the height of the dots is about 0.46±0.20 nm and 0.64±0.25 nm, respectively, i.e. approximately 2-3 ML. Note that these values should correspond to the QD height above the wetting layer.

In the case of high-density (close to 10$^{12}$ cm$^{-3}$) QDs, as the samples presented here, it is difficult to extract reliable measurements of the QD base diameter from AFM. The dimension of the AFM tip is comparable to the dimension of the dots, and deconvolution of the effect of the AFM tip would require precise knowledge of the tip shape. Therefore, to assess the QD geometry, samples SG2, SA1 and SA3 were analysed using cross-section HAADF-STEM and HRTEM, with the results presented in figure 3. The QDs are clearly resolved, with facets that form an angle of ≈ 32° with the (0001) plane. The QD base diameter is 6.9±1.0 nm, 6.2±1.0 nm and 6.5±1.0 nm, and the QD height (including wetting layer) is 5±1 ML, 5.5±1.0 ML and 4±1 ML, for samples SG2, SA1 and SA3, respectively. In addition, the wetting layer thickness is estimated at 2 ML, 2-3 ML and 1-2 ML for SG2, SA1 and SA3, respectively. The average values and error bars were extracted by analysing images of at least 10 QDs per sample. In terms of QD height, the results are consistent with AFM measurements. Comparing the results of SG2 (GaN/AlN QDs) and SA1 ($Al_{0.1}Ga_{0.9}N$/AlN QDs), grown with the same Ga flux, we conclude that the additional Al flux in SA1 does not introduce a significant distortion of the QD morphology. In the case of SA3, the metal fluxes are reduced by approximately a factor of two with respect to SA1, keeping the same Al/Ga ratio. As a result, the QD height reduces drastically, but the diameter of the dots remains approximately constant. This observation points to a temperature-limited QD diameter, i.e. the diameter is given by the adatom diffusion length, which is determined by the substrate temperature.

To study the optical performance of the QDs, the CL emission of the dots at room temperature was recorded and compared as shown in figure 4. The measurements presented here were recorded in an FE-SEM set-up with the accelerating voltage fixed at 5 kV. Solid(dashed) lines represent $Al_{0.1}Ga_{0.9}N$(GaN) QDs, and the spectra are vertically shifted, keeping together samples that were grown with the same gallium flux, decreasing from $\Phi_{Ga}$ = 0.380 ML/s (bottom) to 0.149 ML/s (top). A blue shift is observed in the peak emission wavelengths of $Al_{0.1}Ga_{0.9}N$ dots as compared to GaN dots with the same $\Phi_{Ga}$. The shift corresponds to an average increase of 250 meV in band gap, which is consistent with the incorporation of 10% of Al in the dots. The peak emission wavelength ($\lambda$) of the samples under study and the emission full width at half maximum (FWHM) are summarized in table 1. Note that the emission relative linewidth, defined as FWHM/$\lambda$, remains in the range of 6-11% for all samples.

To evaluate the agreement of the experimental emission wavelength with theoretical expectations, and to analyse the sensitivity of the emission to fluctuations in structural parameters, we have performed 3D simulations of the strain distribution, band diagram and quantum confined levels of SG2, SA1 and SA3, using the structural characterization data as input parameters. The computed values of $\varepsilon_{xx}$, $\varepsilon_{zz}$, $\varepsilon_{zz}/\varepsilon_{xx}$, the band gap, and the resulting $e_1$-$h_1$ transition energy and emission wavelength for the nominal SA1, SG2, and SA3 structures and several structural variations are listed in tables 2, 3, and 4, respectively. Using nominal parameters, i.e. the wetting layer thickness, QD height and base diameter extracted from TEM and the nominal Al composition, we obtain theoretical transition wavelengths of 325±14 nm, 310±7 nm, and 288±13 nm, respectively, in good agreement with experimental values (329 nm, 312 nm, and 270 nm, respectively). The larger deviation in the case of SA3 is due to the fact that the model is reaching its validity limits, since the QDs are only 4 ML high, so that a fluctuation of thickness of 1 ML corresponds to 25% of the height.

Using SA1 as a model structure, figure 5 shows the perturbation of the electronic structure introduced by variations of the QD diameter and height. Figures 5(a) and (b) show the evolution of the electron square wavefunction in a QD located in the centre of the stack, as a function of the QD diameter (nominal and ±1 nm) and the QD height (nominal and ±1 ML). Note that the variations are introduced in such a way that the angle of the QD facets remains constant (≈ 32° with the base, as it corresponds to {10-13} planes). Due to the internal electric field, the electron is shifted towards the apex of the dot, which increases its sensitivity of the lateral confinement. Thus, laterally, the wavefunction spreads when increasing the QD diameter or decreasing the QD height, and it concentrates when reducing the QD diameter or increasing the QD height. Additionally, a vertical displacement of the wavefunction is clearly observed when varying the QD height. The magnitude of the displacement of the electron [0.43 nm between the extreme situations in figure 5(b)] is identical (difference smaller than 0.2%) to the shift observed when varying the thickness of quantum well by the same amount (from 4.5 to 6.5 ML), which points to the fact that the effect of the lateral confinement is relatively weak, in spite of the shape of the electron wavefunction (larger in-plane spread for smaller QDs).

Figure 5(c) represents the hole square wavefunction in the nominal structure, whose shape does not present significant changes when varying the structural parameter. The probability of finding the hole is maximum at the bottom of the QDs. The electric field pushes the hole towards the wetting layer, but it remains laterally confined in the dot. This is because the valence band presents an in-plane maximum at the centre of the QDs, since it is the point with minimum strain. The in-plane ($\varepsilon_{xx}$) and out-of-plane ($\varepsilon_{zz}$) strain, displayed in figure 5(c) for the nominal structure, are maximum in the wetting layer between two dots, whereas they decrease when penetrating the QD. The effect is more remarkable in the case of $\varepsilon_{zz}$. In an AlGaN/AlN quantum well, the compressive stress imposed by the smaller in-plane lattice parameter of AlN results in a compressive (negative) $\varepsilon_{xx}$ in the quantum well. As a reaction to minimize the elastic energy in the layer, the out-of-plane lattice expands following $\varepsilon_{zz} = -(2\, c_{13}/c_{33})\varepsilon_{xx}$, where $c_{13}$ and $c_{13}$ are elastic constants. If we assume $c_{13} = 106$ GPa and $c_{33} = 398$ GPa for GaN [30], $c_{13} = 108$ GPa and $c_{33} = 373$ GPa for AlN [31], and a linear interpolation for the ternary alloy, we obtain $\varepsilon_{zz} = -\left(\frac{2c_{13}}{c_{33}}\right)\varepsilon_{xx} = -0.537\varepsilon_{xx}$ in $Al_{0.1}Ga_{0.9}N$. This describes approximately the behaviour of the wetting layer between the dots. However, in the dots, the presence of AlN in contact with the facets introduces a vertical compressive stress which results in $\varepsilon_{zz}$ values close to zero in figure 5(c).

Let us concentrate now on the evolution of the expected emission wavelength calculated from the energy difference between the first electron and hole levels ($e_1$-$h_1$) and depicted in figures 5(d) as a function of the QD diameter. In spite of the fact that the displacement of the electron wavefunction in figures 5(a) is minimum, the emission energy presents a strong red shift when increasing the QD diameter. The spectral shift of 12.6 nm when changing the diameter by ±1 nm (variation that we had observed in the structural characterization of the sample) is comparable to the emission linewidth (FWHM = 19 nm at room temperature). To understand this shift, it is necessary to look at the evolution of the $\varepsilon_{zz}/\varepsilon_{xx}$ strain ratio, depicted in figure 5(e). In QDs with small diameter, $\varepsilon_{zz}/\varepsilon_{xx}$ differs significantly from what is expected in a planar structure (−0.537), even reaching positive values. As the diameter increases, $\varepsilon_{zz}/\varepsilon_{xx}$ decreases, but it is still as high as −0.075 for a diameter of 7.2 nm. The variation of the strain state has direct impact on the band gap of the material via the band gap deformation potentials, as is illustrated in figure 5(e). Therefore, the shift of the emission wavelength is mostly induced by the modification of the strain state.

If we look at the evolution of the emission wavelength as a function of the QD height, represented in figure 5(f) for various values of QD diameter, an increase in height results in a slight red shift of the emission. However, the effect is moderate, only 2-4 nm when varying the height in the ±1 ML range that marks the structural dispersion, in spite of the clear vertical displacement of the electron wavefunction in figure 5(b). This can also be understood looking at the evolution of $\varepsilon_{zz}/\varepsilon_{xx}$ as a function of the QD height and its effect on the band gap, both depicted in figure 5(g). The increase of $\varepsilon_{zz}/\varepsilon_{xx}$ when increasing the QD height results in a larger band gap, which partially compensates the red shift tendency associated with the smaller vertical confinement and enhanced quantum confined Stark effect. Comparing the predicted shifts in figures 5(d) and 5(f) and the experimental result with the error bars that describe the dot-to-dot structural fluctuations, it becomes clear that the QD diameter plays an important role in the determination of the emission wavelength and linewidth. We have also analysed the effect of varying the Al mole fraction in the dots and wetting layer by ±1% (see table 2), which corresponds to our error bar in the MBE flux calibration for this Al composition. We observe that the magnitude of the spectral shift induced by the deviation in composition would be ±1.8 nm, very small in comparison with the effect of the strain.

The detailed analysis presented here for SA1 is valid for other structures. For instance, looking at the results obtained for SG2, compiled in table 3, fluctuations of the QD diameter by ±1 nm result in a spectral shift of 13.3 nm, larger than the effect of varying the QD height by ±1 ML, which leads to a spectral shift of 11.9 nm. In the extreme case of SA3 (table 4), with QDs that are only 4 ML high, the effect of the diameter is relevant (spectral shift of 6.9 nm for a variation of the diameter by ±1 nm) but smaller than the effect of varying the height by ±1 ML (spectral shift of 16.1 nm), which appears as the major cause of spectral broadening.

Let us now turn to consider the behaviour of these QD superlattices when pumped with an electron beam. To attain high conversion efficiencies, the active region should be able to collect a maximum number of electron-hole pairs generated by the impinging electrons via impact ionization. Therefore, we need to be sure to create the electron-hole pairs within the active region, to ensure minimum loss of energy. To estimate the penetration depth of the beam, we performed Monte Carlo simulations using the CASINO software. Figure 6(a) displays the energy loss of the impinging beam as it penetrates into the structure, assuming that it is pure AlN. We considered various values of acceleration voltage, $V_A$. From the figure, the energy loss associated to the electron-hole generation process takes place within the 530 nm of the active region for $V_A ≤ 7.5$ kV and should hence lead to maximum energy conversion. Note that the fact of considering the structure consisting of pure AlN describes the worst-case scenario, since the density of GaN ($\rho = 6.10 \, g/cm^{-3}$) is almost twice that of AlN ($\rho = 3.26 \, g/cm^{-3}$), and the penetration depth of the electron beam is inversely proportional to the material density [32].

Experimental measurements of CL as a function of the accelerating voltage were also conducted to assess the penetration depth of the electron beam in operating conditions. We studied all the samples at acceleration voltages ranging from 2 kV to 30 kV. As an example, the spectra of SA3, normalized by their maximum and shifted vertically for clarity, are presented in figure 6(b). For low acceleration voltages ($V_A ≤ 15$ kV), we observe a single emission line, assigned to the QDs, at 270 nm. An additional emission at 330 nm appears for $V_A > 20$ kV. This is commonly assumed to be caused by carbon contamination in the AlN template [33–35]. The variation of the QD emission intensity normalized by the injection current (in the range of 500-614 nA) as a function of $V_A$ is displayed in figure 6(c). A saturation of the intensity is observed around $V_A = 9$ kV. Dividing the intensity by the injected power density, we obtain the variation of the emission efficiency as a function of $V_A$ [figure 6(d)]. The efficiency remains stable until $V_A = 9$ kV. Note that the total thickness of the active layers is around 530 nm. Comparing the experimental results with the Monte Carlo simulations in figure 6(a), the electron penetration depth for $V_A$ = 9 kV corresponds approximately to the total thickness of the active layers. Therefore, we can safely say that the efficiency remains approximately constant across the entire active region.

Using the electron gun focused on a 4±1 mm spot, we measured the QD emission as a function of the injection current to assess if our samples could work under high currents without saturation. As an example, the results obtained for SA3 are presented in figure 7(a). We first fixed $V_A$ = 5 kV (which we assume is close to operation conditions, to prevent x-ray emission) and observed no saturation up to 800 µA. In contrast, for $V_A$ = 10 kV we observed a saturation of the emission intensity for injection current higher than 400 µA, presumably due to charging. This could be explained by the fact that, at 10 kV, part of the electron-hole pairs are generated in the AlN template, which makes it more difficult to evacuate the excess electrons. Interestingly, the saturation is associated to a spectral red shift of the emission, as illustrated in figure 7(b). This shift can be partially due to thermal effects, and to the fact that smaller QDs, emitting at shorter wavelengths, are more prone to charging effects than larger QDs, emitting at longer wavelengths.

To quantify the IQE of the structures, temperature-dependent PL measurements were carried out with, first, a continuous wave laser at low power density ($\approx$ 1.3 µW/cm$^2$). The result is illustrated in figures 8(a) and (b), where SG2 is used as an example. We take the liberty to calculate the IQE of the samples as the ratio of the integrated luminescence intensity at room temperature and at low temperature, i.e. IQE $\approx I$(300 K)/$I$(0 K). This equation assumes that at 0 K nonradiative recombination is negligible. The resulting IQE can be overestimated if nonradiative recombination centres are active at low temperature or if nonradiative recombination centres are saturated by high power injection at room temperature. Figure 8(b) shows the integrated PL intensity as a function of the inverse temperature. The PL intensity remains approximately constant until 100 K and then drops exponentially. The trend is well described by the Arrhenius equation: $I(T) \propto 1/(1 + A exp(-E_a/kT))$, where $k$ is Boltzmann constant, and $A$ is a fitting parameter and $E_a$ is the thermal activation energy of the dominant nonradiative process ($E_a$ = 62±4 meV in the figure). This behaviour supports our assumption that nonradiative processes are negligible at low temperature, and the low power density of the laser (low injection) allows measuring the intrinsic IQE of the structures, without saturation of nonradiative centres. The values of the IQE measured this way are listed in table 1, together with $E_a$ for various samples. IQE values around 50% are systematically obtained for QDs grown with $\Phi_{metal}/\Phi_N$ < 0.75. This is an improvement with respect to previous reports, where similar characterization (low injection) rendered IQE values in the range of 10-40% for GaN/AlN QDs grown with $\Phi_{metal}/\Phi_N$ = 0.8-0.9 [18,23]. If we compare with the results obtained by Himwas et al. for $Al_{0.1}Ga_{0.9}N$/AlN QDs with $\Phi_{metal}/\Phi_N$ = 0.63 [22], IQEs around 50% were only achieved for QD layers generated from 4-5 ML of $Al_{0.1}Ga_{0.9}N$. Our results here point to the fact that further reduction of the metal-to-nitrogen ratio increases the efficiency of small QDs, so that IQE around 50% is possible for QDs generated from only 3 ML of either GaN or $Al_{0.1}Ga_{0.9}N$.

It is however true that most IQE values reported in literature are measured with a pulsed laser and at much higher power densities [14,21,36–44], in the range of 5-1000 kW/cm² to be close to the operation conditions of LEDs. This procedure leads to higher IQE values, which depend on the pulse width, repetition rate and power density of the pumping laser [36–39,41,45]. Nevertheless, our samples have been characterized with a pulsed laser (266 nm, 2 ns pulses, repetition rate of 8 kHz) and the IQE variation as a function of the excitation power density was recorded purely for the sake of comparison with literature. Measurements were performed at 6 K and at 300 K. Under high excitation, the calculation of the IQE at room temperature must take into account the drop of the PL efficiency at low temperature due to the many-body effects induced by high-power excitation [41,46] so that

$$IQE(300K, P) = \frac{I(300K,P)}{I(0K,P)} \times \frac{I(0K,P)/P}{I(0K,P_{li})/P_{li}} \quad (1)$$

where $I(T,P)$ is the integrated PL intensity as a function of temperature and excitation power ($P$), and $P_{li}$ is an excitation power at low excitation conditions. The results are displayed in Figure 8(c). The GaN/AlN and $Al_{0.1}Ga_{0.9}N$/AlN SK QD samples (SG2 and SA1 in the figure, respectively) are compared with GaN/AlN QDs in nanowires and GaN/AlN multi-quantum-wells (from ref. [46]). The efficiency of both SK QDs and nanowires remain quite stable as a function of the pumping power. In the case of SK QDs, the IQE remains approximately constant for pumping power densities as high as 200 kW/cm². This proves our QDs are by far superior to other heterostructures in terms of stability under high excitation. However, it should be kept in mind that, in an electron-pumped UV lamp using an acceleration voltage of 5 kV and an injection current of 1 mA to irradiate a spot with a diameter of 1 mm, the excitation density would be below 1 kW/cm². This situation is clearly unfavourable for the use of quantum well structures.

## 5. Conclusion

To summarize, we have demonstrated 100-period $Al_xGa_{1-x}N$/AlN (0 ≤ x ≤ 0.1) QD superlattices emitting in the 244-335 nm range at room temperature, with a relative linewidth in the 6-11% range. The $Al_xGa_{1-x}N$/AlN active region is 530 nm long (with 100 periods of QDs), which is enough to collect the electron-hole pairs generated by an electron beam with an acceleration voltage $V_A \leq 9$ kV. This was experimentally confirmed with extensive CL studies. There was no significant variation of the efficiency or emission line width for acceleration voltages in the $2 \leq V_A \leq 9$ kV range, which indicates that the nanostructures are homogeneous along the growth axis, and the linewidth is mostly limited by the in-plane fluctuations of the QD diameter. When measured at 5 kV, the UV emission of the heterostructure scales linearly with the injected current in the measured range up to 800 µA (with continuous pumping using an electron gun). For $V_A = 10$ kV, a saturation is observed for injection currents higher than 400 µA which is attributed to enhanced charging effects due to the injection of carriers in the AlN substrate. Then, the IQEs of all the samples were quantified, obtaining values around 50% for QDs synthesized with III/V ratio < 0.75. Efficiencies remained stable as a function of the pumping power up to high as 200 kW/cm$^2$, proving that SK QDs can be used for application in devices with various pumping requirements.

## Acknowledgements

This work is supported by the French National Research Agency (ANR) via the UVLASE program (ANR-18-CE24-0014), and by the Auvergne-Rhône-Alpes region (grant PEAPLE). We acknowledge technical support from F. Jourdan, Y. Curé and Y. Genuist. We benefited from the access to the technological platform NanoCarac of CEA-Minatech Grenoble in collaboration with the IRIG-LEMMA group.

**Table 1.** Growth parameters and optical characteristics of the samples under study: Ga and Al fluxes in monolayers per second ($\Phi_{Ga}$ and $\Phi_{Al}$, respectively), metal-to-nitrogen ratio ($\Phi_{metal}/\Phi_N$), number of monolayers deposited for the generation of each QD layer, nominal Al concentration in the dots (x), peak emission wavelength at room temperature (in the case of multiple peaks, the dominant peak appears in **bold** fonts), full width at half maximum (FWHM) of the main emission line at room temperature, internal quantum efficiency (IQE) at room temperature measurement under low-injection conditions, and activation energy ($E_a$) of the main nonradiative process extracted from variation of the PL intensity as a function of temperature.

| Sample | $\Phi_{Ga}$ (ML/s) | $\Phi_{Al}$ (ML/s) | $\Phi_{metal}/\Phi_N$ | Number of MLs | x | Peak emission (nm) | FWHM (nm) | IQE (%) | $E_a$ (meV) |
|---|---|---|---|---|---|---|---|---|---|
| SG1 | 0.441 | 0 | 0.85 | 5.3 | 0 | 335 | 24 | 35 | 55±8 |
| SG2 | 0.380 | 0 | 0.73 | 4.6 | 0 | 329 | 28 | 50 | 65±6 |
| SG3 | 0.319 | 0 | 0.61 | 3.8 | 0 | 310 | 35 | 51 | 66±5 |
| SG4 | 0.306 | 0 | 0.59 | 3.7 | 0 | 316 | 18 | 61 | 74±7 |
| SG5 | 0.220 | 0 | 0.42 | 2.6 | 0 | 294 | 32 | 50 | 46±5 |
| SG6 | 0.211 | 0 | 0.41 | 2.5 | 0 | 294 | 28 | 50 | 49±7 |
| SG7 | 0.149 | 0 | 0.29 | 1.8 | 0 | 250 | 16 | 43 | 100±40 |
| SA1 | 0.380 | 0.038 | 0.80 | 5.0 | 0.1 | 312 | 19 | 52 | 78±13 |
| SA2 | 0.319 | 0.032 | 0.68 | 4.2 | 0.1 | 298 | 21 | 54 | 70±10 |
| SA3 | 0.220 | 0.022 | 0.47 | 2.9 | 0.1 | 270 | 22 | 47 | 68±7 |
| SA4 | 0.149 | 0.015 | 0.32 | 2.0 | 0.1 | **244**, 277 | 15 | 33 | 68±14 |

**Table 2.** Simulations based on the structure of SA1 (experimental emission wavelength = 312 nm). Input structural parameters include QD height, QD base diameter, wetting layer (WL) thickness and Al mole fraction (x) in the dots and in the WL. Nominal values are marked in **bold**. As relevant output data, we collect here the in-plane and out-of-plane strain ($\varepsilon_{xx}$ and $\varepsilon_{zz}$, respectively) at the point of the maximum of the electron wavefunction, as well as the $\varepsilon_{zz}/\varepsilon_{xx}$ ratio and the resulting band gap ($E_G$) at the same point, the energy difference between the first electron level ($e_1$) and the first hole level ($h_1$), and the corresponding expected emission wavelength ($\lambda$).

| Simulation | QD height (ML) | QD diameter (nm) | WL thickness (ML) | x | $\varepsilon_{xx}$ (%) | $\varepsilon_{zz}$ (%) | $\varepsilon_{zz}/\varepsilon_{xx}$ | $E_G$ (eV) | $e_1$-$h_1$ (eV) | $\lambda$ (nm) |
|---|---|---|---|---|---|---|---|---|---|---|
| **Nominal** | **5.5** | **6.2** | **2.5** | **0.10** | -1.16 | -0.016 | 0.010 | 3.944 | **4.001** | **309.9** |
| H-1 | 4.5 | **6.2** | **2.5** | **0.10** | -1.75 | 0.30 | -0.170 | 3.930 | 4.041 | 306.8 |
| H+1 | 6.5 | **6.2** | **2.5** | **0.10** | -1.44 | -0.36 | 0.247 | 3.960 | 3.992 | 310.6 |
| D-1 | **5.5** | 5.2 | **2.5** | **0.10** | -1.45 | -0.23 | 0.159 | 3.954 | 4.091 | 303.1 |
| D+1 | **5.5** | 7.2 | **2.5** | **0.10** | -1.67 | 0.12 | -0.075 | 3.939 | 3.928 | 315.7 |
| H-1D-1 | 4.5 | 5.2 | **2.5** | **0.10** | -1.67 | 0.14 | -0.086 | 3.936 | 4.113 | 301.5 |
| H+1D-1 | 6.5 | 5.2 | **2.5** | **0.10** | -1.35 | -0.58 | 0.432 | 3.973 | 4.091 | 303.1 |
| H-1D+1 | 4.5 | 7.2 | **2.5** | **0.10** | -1.79 | 0.39 | -0.217 | 3.927 | 4.013 | 309.0 |
| H+1D+1 | 6.5 | 7.2 | **2.5** | **0.10** | -1.54 | -0.16 | 0.106 | 3.952 | 3.903 | 317.7 |
| Al-1 | **5.5** | **6.2** | **2.5** | 0.09 | -1.61 | -0.016 | 0.010 | 3.919 | 3.978 | 311.7 |
| Al+1 | **5.5** | **6.2** | **2.5** | 0.11 | -1.58 | -0.015 | 0.009 | 3.970 | 4.025 | 308.1 |

**Table 3.** Simulations based on the structure of SG2 (experimental emission wavelength = 329 nm). Input structural parameters include QD height, QD base diameter, wetting layer (WL) thickness and Al mole fraction (x) in the dots and in the WL. Nominal values are marked in **bold**. As relevant output data, we collect here the in-plane and out-of-plane strain ($\varepsilon_{xx}$ and $\varepsilon_{zz}$, respectively) at the point of the maximum of the electron wavefunction, as well as the $\varepsilon_{zz}/\varepsilon_{xx}$ ratio and the resulting band gap ($E_G$) at the same point, the energy difference between the first electron level ($e_1$) and the first hole level ($h_1$), and the corresponding expected emission wavelength ($\lambda$).

| Simulation | QD height (ML) | QD diameter (nm) | WL thickness (ML) | x | $\varepsilon_{xx}$ (%) | $\varepsilon_{zz}$ (%) | $\varepsilon_{zz}/\varepsilon_{xx}$ | $E_G$ (eV) | $e_1$-$h_1$ (eV) | $\lambda$ (nm) |
|---|---|---|---|---|---|---|---|---|---|---|
| **Nominal** | **5.0** | **6.9** | **2.0** | **0** | -1.88 | 0.15 | -0.078 | 3.678 | 3.818 | 324.8 |
| H-1 | 4.0 | **6.9** | **2.0** | **0** | -2.03 | 0.52 | -0.256 | 3.663 | 3.912 | 317.0 |
| H+1 | 6.0 | **6.9** | **2.0** | **0** | -1.74 | -0.10 | 0.057 | 3.691 | 3.770 | 328.9 |
| D-1 | **5.0** | 5.9 | **2.0** | **0** | -1.81 | 0.07 | -0.042 | 3.684 | 3.928 | 315.7 |
| D+1 | **5.0** | 7.9 | **2.0** | **0** | -1.93 | 0.25 | -0.129 | 3.674 | 3.769 | 329.0 |
| H-1D-1 | 4.0 | 5.9 | **2.0** | **0** | -1.98 | 0.41 | -0.209 | 3.667 | 4.005 | 309.6 |
| H+1D+1 | 6.0 | 7.9 | **2.0** | **0** | -1.18 | 0.50 | -0.273 | 3.685 | 3.730 | 332.4 |

**Table 4.** Simulations based on the structure of SA3 (experimental emission wavelength = 270 nm). Input structural parameters include QD height, QD base diameter, wetting layer (WL) thickness and Al mole fraction (x) in the dots and in the WL. Nominal values are marked in **bold**. As relevant output data, we collect here the in-plane and out-of-plane strain ($\varepsilon_{xx}$ and $\varepsilon_{zz}$, respectively) at the point of the maximum of the electron wavefunction, as well as the $\varepsilon_{zz}/\varepsilon_{xx}$ ratio and the resulting band gap ($E_G$) at the same point, the energy difference between the first electron level ($e_1$) and the first hole level ($h_1$), and the corresponding expected emission wavelength ($\lambda$).

| Simulation | QD height (ML) | QD diameter (nm) | WL thickness (ML) | x | $\varepsilon_{xx}$ (%) | $\varepsilon_{zz}$ (%) | $\varepsilon_{zz}/\varepsilon_{xx}$ | $E_G$ (eV) | $e_1$-$h_1$ (eV) | $\lambda$ (nm) |
|---|---|---|---|---|---|---|---|---|---|---|
| **Nominal** | **4.0** | **6.5** | **1.5** | **0.1** | **-1.80** | **0.42** | **-0.232** | **3.925** | **4.307** | **287.9** |
| H-1 | 3.0 | **6.5** | **1.5** | **0.1** | -1.9 | 0.66 | -0.345 | 3.914 | 4.461 | 277.9 |
| H+1 | 5.0 | **6.5** | **1.5** | **0.1** | -1.7 | 0.13 | -0.08 | 3.937 | 4.217 | 294.0 |
| D-1 | **4.0** | 5.5 | **1.5** | **0.1** | -1.7 | 0.288 | -0.165 | 3.93 | 4.370 | 283.8 |
| D+1 | **4.0** | 7.5 | **1.5** | **0.1** | -1.8 | 0.5 | -0.27 | 3.921 | 4.266 | 290.7 |
| H-1D-1 | 3.0 | 5.5 | **1.5** | **0.1** | -1.8 | 0.589 | -0.312 | 3.917 | 4.505 | 275.3 |
| H+1D+1 | 5.0 | 7.5 | **1.5** | **0.1** | -1.7 | 0.251 | -0.145 | 3.933 | 4.152 | 298.7 |
| Al+1 | **4.0** | **6.5** | **1.5** | 0.11 | -1.7 | 0.414 | -0.232 | 3.952 | 4.33 | 286.4 |

**Figure Captions**

**Figure 1.** (a) Schematic description of the QD superlattices under study. (b) Reciprocal space map around the (10-15) reflection of sample SG5. Labels indicate the reflection of AlN and the main reflection of the QD superlattice (SL), which appear vertically aligned. (c) XRD θ–2θ scans of samples SA1, SA2 and SA3 recorded around the (0002) reflection of AlN. The scans are vertically shifted for clarity. Labels indicate the (0002) reflection of AlN and the QD superlattice (SL) with several satellites. The peak around 41.7° corresponds to the (0006) reflection of the sapphire template.

**Figure 2.** AFM images of samples SG3 and SA1.

**Figure 3.** Top: HAADF-STEM images of samples (a) SG2, (b) SA1 and (c) SA3 showing 6-7 periods of the GaN/AlN quantum dot superlattice. Dark/bright contrast corresponds to Al-rich/Ga-rich areas. Bottom: HRTEM off-axis (10° tilt from the [11–20] zone axis) images of the same samples. Here bright/dark contrast corresponds to Al-rich/Ga-rich areas. The colour code is associated with the sample and is consistent along the paper.

**Figure 4.** Room-temperature CL emission from GaN/AlN (dashed) and $Al_{0.1}Ga_{0.9}N$/AlN quantum dot superlattices grown with different fluxes of gallium (from bottom to top, $\Phi_{Ga}$ = 0.380, 0.319, 0.220, and 0.149 ML/s). The spectra are normalized to their maxima and vertically shifted for clarity.

**Figure 5.** Calculations based on the structural parameters of SA1 (Nominally QDs with height = 5.5 ML, base diameter = 6.2 nm, wetting layer height = 2.5 ML, Al mole fraction in the QDs and in the wetting layer = 0.1). (a) Evolution of the electron square wavefunction ($e_1$) as a function of the QD diameter (5.2 nm, 6.2 nm, and 7.2 nm). The calculation with nominal parameters is framed in orange. (b) Evolution of the electron square

wavefunction ($e_1$) as a function of the QD height (4.5 ML, 5.5 ML, and 6.5 ML). (c) For the nominal structure, hole square wavefunction ($h_1$), in-plane strain ($\varepsilon_{xx}$) and out of plane strain ($\varepsilon_{zz}$). (d) Variation of the emission wavelength as a function of the QD diameter. The experimental data with its structural error bar appears as an orange diamond. (e) Variation of the $\varepsilon_{zz}/\varepsilon_{xx}$ ratio and the band gap ($E_G$) at the maximum of the electron wavefunction, as a function of the QD diameter. (f) Variation of the emission wavelength as a function of the QD height for various values of QD diameter. The experimental data with its structural error bar appears as an orange diamond. (g) Variation of the $\varepsilon_{zz}/\varepsilon_{xx}$ ratio and the band gap ($E_G$) at the maximum of the electron wavefunction, as a function of the QD height. The results are given for various values of QD diameter.

**Figure 6.** (a) Energy loss as a function of depth by an electron beam penetrating in AlN for various accelerating voltages, $V_A$. The curves were obtained by performing Monte Carlo simulations using the CASINO software. (b) CL measurements of sample SA3 as a function of the acceleration voltage, $V_A$. Spectra are normalized at the maximum intensity value and vertically shifted for clarity. The emission from the quantum dot superlattice peaks at 270 nm. The band around 330 nm that appears for $V_A > 20$ kV is assigned to carbon contamination in the AlN template. (c) Normalized CL intensity and (d) emission efficiency as a function of $V_A$ for the same sample. Measurements were performed using an electron gun operated in direct current mode, under normal incidence, with a beam spot diameter of 4±1 mm.

**Figure 7.** (a) Variation of CL intensity as a function of the injection current measured for SA3 at $V_A = 5$ kV and 10 kV. The slope of the solid grey line corresponds to a linear increase. No saturation is observed up until 800 μA for $V_A = 5$ kV while a saturation for currents higher than 400 μA is observed at $V_A$ = 10 kV. (b) Spectral shift observed as a function of the injection current.

**Figure 8.** (a) PL spectra from sample SG2 measured at various temperatures under low excitation density conditions (100 μW laser power, focused on a spot with a diameter of 100 μm). (b) In the same sample, variation of the integrated PL intensity as a function of the inverse temperature. The dashed line is a fit to $I(T) \propto 1/(1 + A exp(-E_a/kT))$, where k is Boltzmann constant, and $A$ and $E_a$ are fitting parameters. An activation energy $E_a$ = 62±4 meV is extracted from the fit. (c) Variation of the IQE at room temperature as a function of the excitation power density measured with a pulsed Nd-YAG laser.

**Figure 1**

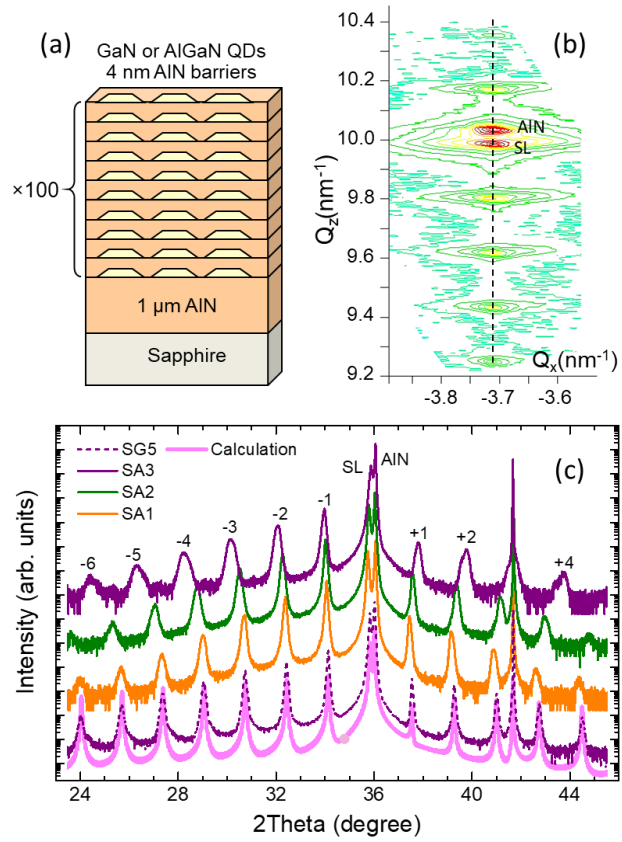

**Figure 2**

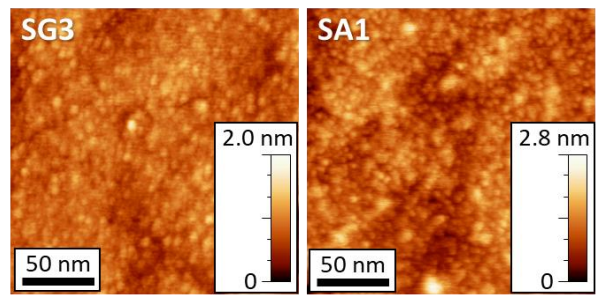

**Figure 3**

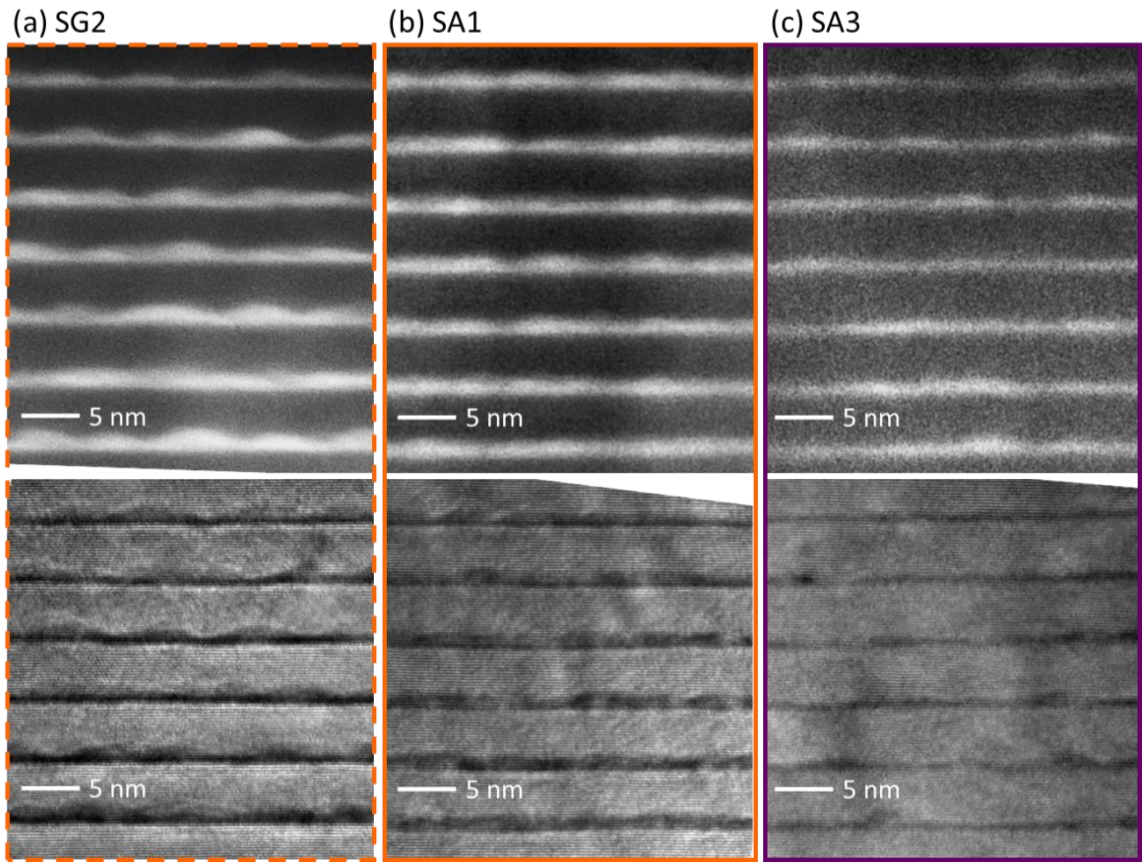

**Figure 4**

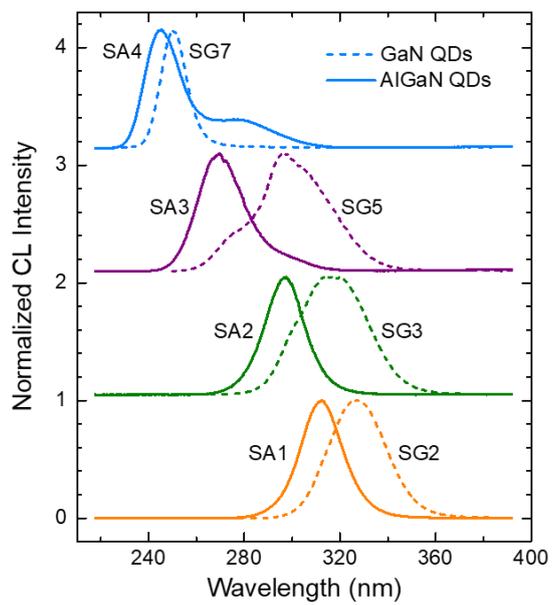

**Figure 5**

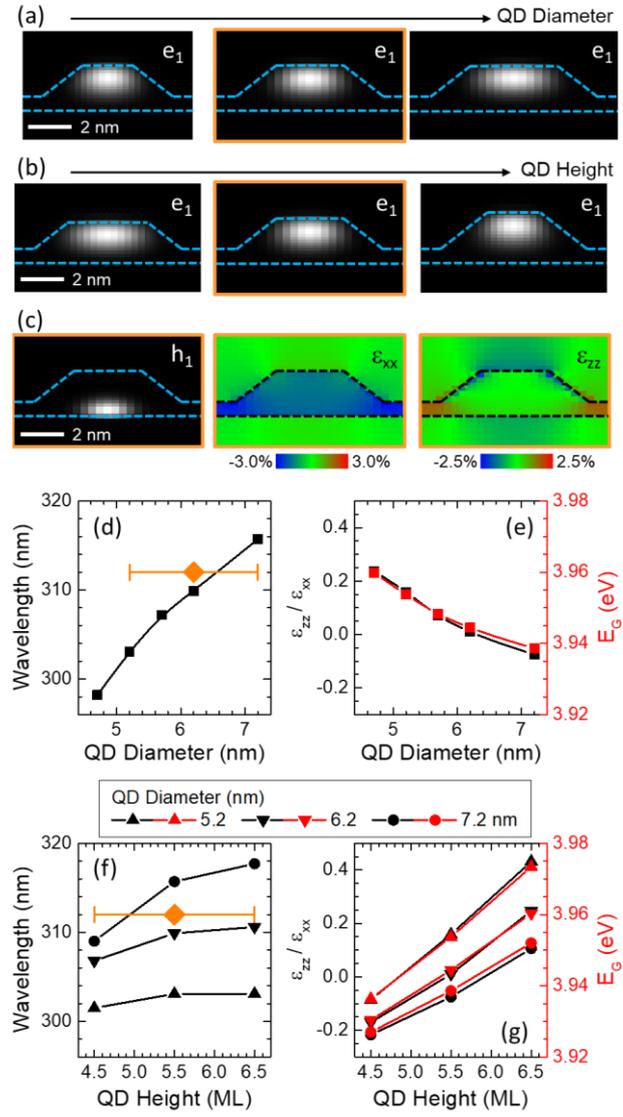

**Figure 6**

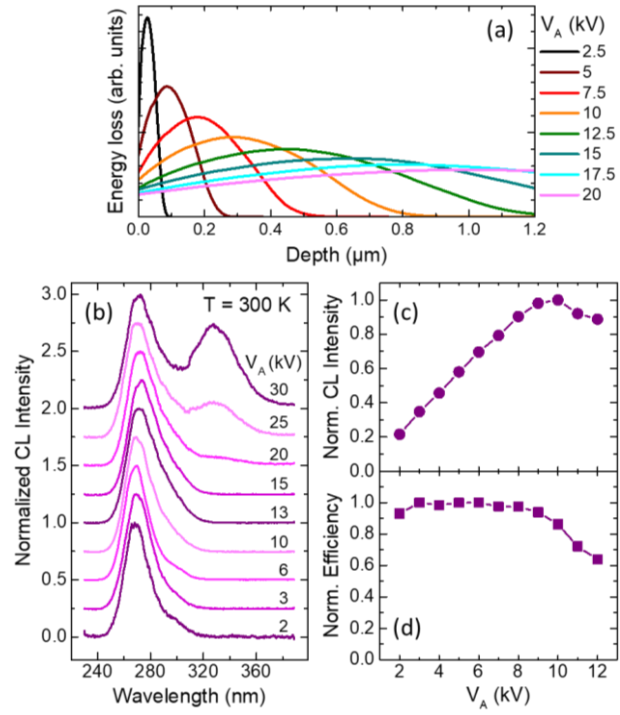

**Figure 7**

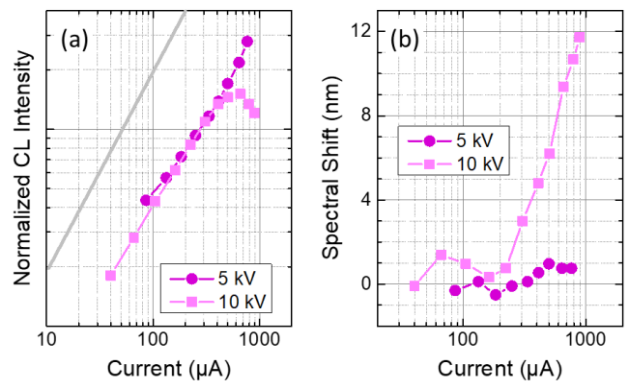

**Figure 8**

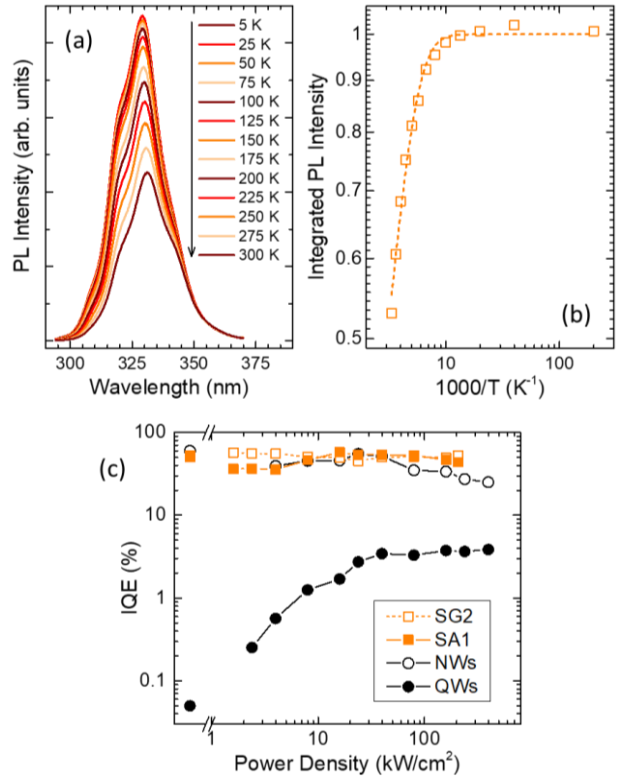